Rainer Burghardt

# Kerr Interior Surfaces

A recently found interior for the Kerr metric is re-investigated by means of geometrical methods. A surface with nonholonomicity is matched to the surface of the exterior solution.

## 1. Introduction

In a former paper [1] we have proposed a solution for a Kerr interior based on a differentially rotating fluid source. In the present paper we explain this solution by geometrical means. In Sec. 2 we choose for the [r, $\vartheta$]-slice of the space-like part of the metric a surface of ellipsoidal shape embedded in a flat space with an extra dimension. This surface matches the surface of the exterior solution (ES), which have we investigated in an earlier paper [2]. The [r, $\varphi$]-slice will not be discussed as it can be obtained in the same way as the [r, $\vartheta$]-slice by reducing the surface of an elliptical shape to that of a spherical one. To include also the time surface five dimensions but six variables and more algebra is needed. This is performed in Sec. 3.

## 2. The space-like geometry

In this Section we investigate the geometrical structure of the space-like part of the Kerr interior. It is sufficient to consider the [r, $\vartheta$]-slice as the [r, $\varphi$]-slice has a similar, even simpler, structure. The [r, $\vartheta$]-slice of the complete Schwarzschild solution is made up of Flamm's paraboloid (a fourth-order surface) and of the cap of a sphere for the interior, covering the 'hole' of the ES. As we have found for the [r, $\vartheta$]-slice of the ES an elliptically squashed surface [2] which reduces to Flamm's paraboloid by setting the rotational parameter to zero, we expect that the interior surface should reduce to a cap of a sphere for the Schwarzschild case. Since the parallels of the ES are ellipses, we demand the parallels of the interior to be elliptical too. Thus, we try a cap of an elliptically squashed surface for the [r, $\vartheta$]-slice and a cap of a sphere for the [r, φ]-slice embedded in a flat space with the extra dimension $x^{0'}$. We can use the elliptic-hyperbolical Boyer-Lindquist co-ordinate system for both solutions as well. The 3-surface is parametrized by

$$x^{0'} = \mathbf{R}\cos\eta$$
$$x^{1'} = \mathbf{R}\sin\eta\cos\vartheta$$
$$x^{2'} = {}^*\mathbf{R}\sin\eta\sin\vartheta\cos\varphi$$
$$x^{3'} = {}^*\mathbf{R}\sin\eta\sin\vartheta\sin\varphi$$

$\hspace{6cm}$ (2.1)

where $\mathbf{R}$ is a constant and the primed indices refer to a Cartesian co-ordinate system in the flat embedding space and

$$\sin\eta = \frac{r}{\mathbf{R}} = \frac{A}{{}^*\mathbf{R}}, \quad \cos\eta = \sqrt{1 - \frac{r^2}{\mathbf{R}^2}}, \quad A = \sqrt{r^2 + a^2}, \quad {}^*\mathbf{R} = \frac{A}{r}\mathbf{R} \cdot \quad (2.2)$$

The horizontals of the surface are confocal ellipses with the minor semi-axes r and the major semi-axes A, where a is the common eccentricity of the ellipses. (2.1) can be written as

$$x^{0'} = \pm\sqrt{\mathbf{R}^2 - r^2}$$
$$x^{1'} = r\cos\vartheta$$
$$x^{2'} = A\sin\vartheta\cos\varphi$$
$$x^{3'} = A\sin\vartheta\sin\varphi$$

$\hspace{6cm}$ (2.3)

The lower half of the resulting surface is shown in Fig. 1.

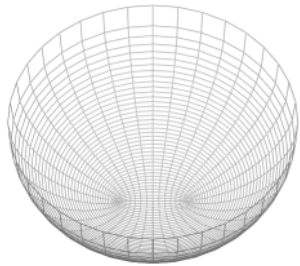

*Fig. 1*

**R** has the meaning of the radius of a circle at the minor semi-axes of the horizontal ellipses. For the space-like part of the Kerr interior we will take a cap of this surface and match it to the ES. Differentiating (2.1) we obtain for the space-like line element of the surface

$$ds^2 = \left[\tan^2 \eta + a_R^2 \right] \mathbf{R}^2 \cos^2\eta \, d\eta^2 + \Lambda^2 d\vartheta^2 + \sigma^2 d\varphi^2 \tag{2.4}$$

with

$$a_R^2 = \frac{\Lambda^2}{A^2} = 1 - \omega^2\sigma^2, \quad \Lambda^2 = r^2 + a^2\cos^2\vartheta = A^2 - a^2\sin^2\vartheta, \quad \omega = \frac{a}{A^2}, \quad \sigma = A\sin\vartheta \,. \tag{2.5}$$

As

$$dr = \mathbf{R}\cos\eta \, d\eta, \quad \alpha_R = 1/a_R \tag{2.6}$$

we obtain the radial part of the line element

$$\sqrt{1 + \alpha_R^2 \tan^2 \eta} \; a_R dr$$

whereas $a_R dr$ is the line element of the BL hyperbolae defined by $\vartheta$=const. in the flat zero-plane $x^{0'} = 0$. Defining the slope of the radial curves on the surface as

$$\tan\beta = \alpha_R (r, \vartheta)\tan\eta \tag{2.7}$$

we get

$$dx^1 = \frac{1}{\cos\beta} a_R dr \,,$$

where the unprimed indices refer to the local reference system, $dx^1$ being tangent to the surface. Finally we get

$$ds^2 = \frac{1}{\cos^2\beta} a_R^2 dr^2 + \Lambda^2 d\vartheta^2 + \sigma^2 d\varphi^2. \tag{2.8}$$

Evidently, the metric (2.8) is not the metric we have proposed in our previous paper and the metric does not match the ES. The cause is the following: transporting the normal vector of this surface from the minor semi-axes of the elliptical horizontals around the $x^{0'}$- axis, this vector will cyclically move up and down because the slope of the surface depends on $\vartheta$, as can be seen from (2.7). From a new rigging vector we demand the off-axis angle to remain constant on its way around. The hyperplanes normal to this vector are anholonomic and the world we are living in is the family of all these hyperplanes. From (2.1) and (2.6) we get

$$dx^{0'}_{holonomic} = -\tan\eta\,dr = -\tan\beta\,a_R dr, \qquad \tan\eta = \mp\frac{r}{\sqrt{\mathbf{R}^2 - r^2}} \qquad (2.9)$$

and we define the non-integrable function

$$dx^{0'}_{anholonomic} = -\tan\eta\,a_R(r,\vartheta)dr \quad . \qquad (2.10)$$

Suppressing the other dimensions, we have the flat radial line element in BL co-ordinates $dx^{1'} = a_R dr$ and obtain the anholonomic radial line element by

$$dx^{0'2} + dx^{1'2} = \frac{1}{\cos^2\eta}a_R^2 dr^2 = \frac{1}{1 - \dfrac{r^2}{\mathbf{R}^2}}a_R^2 dr^2 \quad . \qquad (2.11)$$

We remark that the holonomic radial line element and the anholonomic radial line element have the same projections $dx^{1'} = a_R dr$ on the zero-plane.

We select the lower part $x^{0'}_{holonomic} = -\sqrt{\mathbf{R}^2 - r^2}$ of the surface and we fix its center by addition of a constant in a suitable way for a proper matching. If this matching excludes the ergosphere we obtain a complete solution for a rotating object avoiding all singularities except the singularity at the rim of a disk in the equatorial plane for r = 0. The boundary value $\eta_g$ is the aperture angle of the cap and $\tan\eta_g$ the slope of the cap at the minor semi-axes of the horizontals. Fig. 2 shows the matching region of the interior and exterior surfaces.

Now we adjust the sign of $\eta$, so that the orientation is cw. Then $\eta_g$ coincides with $\varepsilon_g$, the angle of ascent of the exterior surface and the interior metric matches the exterior one at the boundary surface. This has the advantage that we do not need to correct the signs of the physical quantities. Thus we get for the IS and ES with

$$\sin\eta = -\frac{r}{\mathbf{R}}, \qquad \sin\varepsilon = v_s = -\frac{r}{A}\sqrt{\frac{2M}{r}} \qquad (2.12)$$

a negative value for the velocity $v_S$ of a freely falling observer (also free from dragging effects) and a negative value for the attractive force of gravity.

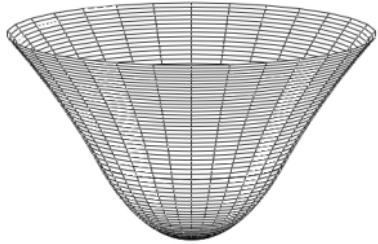

*Fig. 2*

## 3. The time surface

In previous papers [3-6] we have shown that it is possible to embed the exterior Schwarzschild solution and the Kerr solution in a five dimensional flat space, if we use six variables without violating the theorems of Kasner and Eisenhart. This has been established by using the theory of double surfaces developed in these papers. With the help of this method we have explained the geometrical background of the Schwarzschild interior solution and we will proceed in the same way to investigate the Kerr interior metric. We start with the static seed metric investigated in the previous paper. For the explanation of the time-like part of the metric the curvature of the radial lines play an important role: We move the radial curvature vector of the ES from infinity towards the stellar object. While the tip of this vector moves on a radial integral line of the exterior surface the tail moves on the correlated evolute. As soon as the tip has reached the boundary surface the tail is held on the evolute and the tip is moved on the interior surface towards the axis of rotation. This new vector we call X and $\varepsilon$ its negative off-axis angle with respect to $x^{0'}$. At the boundary surface X coincides with the curvature vector $\rho_g$ of the radial lines of the ES. If we prolong $\rho_g$ to $x^{0'}$ the straight line through $\rho_g$ cuts off a distance

$$\overline{\mathbf{R}} = \mathbf{R} + \rho_g \ . \tag{3.1}$$

The components of the vector X with respect to the Cartesian co-ordinate system of the embedding space are

$$X^{0'} = X \cos \varepsilon = \overline{\mathbf{R}} \cos \eta_g - \mathbf{R} \cos \eta$$
$$X^{1'} = X \sin \varepsilon = \overline{\mathbf{R}} \sin \eta_g - \mathbf{R} \sin \eta \qquad (3.2)$$

We introduce an additional dimension with the co-ordinate axis $x^{4'}$ and we define the time as the rotation of the projection of the vector X onto the $[x^{0'}, x^{4'}]$-plane through the imaginary angle $i\psi$:

$$d\, x^4 = X \cos \varepsilon di\psi = \left[ \overline{\mathbf{R}} \cos \eta_g - \mathbf{R} \cos \eta \right] d\, i\psi \cdot \qquad (3.3)$$

Since the value of the curvature vector of the exterior curve at the minor semi-axis is known from previous investigations [5] as

$$\rho = -\frac{2r}{\sin \varepsilon} \frac{r^2 + a^2}{r^2 - a^2}\,, \qquad (3.4)$$

we have at the boundary

$$\rho_g = -\frac{2\, r_g}{\sin \eta_g} \Phi_g^2, \quad \Phi_g^2 = \frac{r_g^2 + a^2}{r_g^2 - a^2} \qquad (3.5)$$

and with (2.12)

$$\rho_g = 2\mathbf{R}\Phi_g^2 \cdot \qquad (3.6)$$

Defining the co-ordinate time as $dt = \rho_g d\psi$ we obtain with (3.1) and (3.3) the physical time in the local tetrad system

$$dx^4 = \frac{1}{2}\left[ \left(1 + 2\Phi_g^2\right) \cos \eta_g - \cos \eta \right] \left(\Phi_g^2\right)^{-2} idt \qquad (3.7)$$

and the seed metric of the previous paper. The transformation to the rotating metric is straight forward. For the Schwarzschild case we obtain $\Phi_g = 1$ and

$$dx^4 = \frac{1}{2}\left[ 3\cos \eta_g - \cos \eta \right] idt = \left[ 3\mathbf{R} \cos \eta_g - \mathbf{R} \cos \eta \right] d\, i\psi \qquad (3.8)$$

elucidating that the curvature vector of the Schwarzschild parabola at the boundary is $\rho_g = 2\mathbf{R}$ and $\overline{\mathbf{R}} = \rho_g + \mathbf{R} = 3\mathbf{R}$, which is a fundamental property of the parabola.

## 4.   More on the geometry

In the last section we have envisaged the vector X in the $[x^{0'}, x^{1'}]$-plane only. Extended to more dimensions we can read from

$$X^{3'} = X \sin \varepsilon \sin \theta \sin \varphi$$
$$X^{2'} = X \sin \varepsilon \sin \theta \cos \varphi$$
$$X^{1'} = X \sin \varepsilon \cos \theta$$
$$X^{0'} = X \cos \varepsilon \cos i\psi$$
$$X^{4'} = X \cos \varepsilon \sin i\psi$$

(4.1)

that X satisfies the equation of a pseudo-hypersphere with radius X

$$X_{a'} X^{a'} = X^2$$

(4.2)

embedded in a five-dimensional flat space with Cartesian co-ordinates $a' = \{0', 1', ..., 4'\}$. The sphere (4.1) provides the basic framework for several solutions of the Einstein field equations with spherical and also axial symmetry. To specify such a model, the sphere has to be deformed to another more complicated surface. This is easily done by projectors expanding the center of the sphere to a curve, which is the locus of the centers of curvatures of another curve. The latter is expanded from a great circle of the sphere. These two correlated curves (evolvente and evolute) are rotated through the above-mentioned angles and constitute a double surface embedded in a five-dimensional flat space [3-6]. By the introduction of a second surface into the theory we can do with five dimensions only for the embedding space for vacuum solutions without contradicting the theorem of Kasner and Eisenhart. The evolute provides a hidden variable and our theory is based on five dimensions but six variables. A dimensional reduction cuts off all that we do not need for the four-dimensional representation of the model.

From (4.1) we can easily derive the transformation matrix to pseudo-spherical co-ordinates $a = \{0, 1, ..., 4\}$ with the co-ordinate labels $\{X, \eta, \theta, \varphi, \psi\}$. $\theta$ is the off-axis angle of the curvature vectors of the BL-ellipses and

$$\sin \theta = \frac{r}{\Lambda} \sin \vartheta, \quad \cos \theta = \frac{A}{\Lambda} \cos \vartheta \cdot$$

With the help of (3.2) we get

$$X^{3'} = \bar{\mathbf{R}}^{3'} - \mathbf{R}^{3'} = \bar{\mathbf{R}} \sin \eta_g \sin \theta \sin \varphi - \mathbf{R} \sin \eta \sin \theta \sin \varphi$$

$$X^{2'} = \bar{\mathbf{R}}^{2'} - \mathbf{R}^{2'} = \bar{\mathbf{R}} \sin \eta_g \sin \theta \cos \varphi - \mathbf{R} \sin \eta \sin \theta \cos \varphi$$

$$X^{1'} = \bar{\mathbf{R}}^{1'} - \mathbf{R}^{1'} = \bar{\mathbf{R}} \sin \eta_g \cos \theta \qquad - \mathbf{R} \sin \eta \cos \theta \qquad . \qquad (4.3)$$

$$X^{0'} = \bar{\mathbf{R}}^{0'} - \mathbf{R}^{0'} = \bar{\mathbf{R}} \cos \eta_g \cos i\psi \qquad - \mathbf{R} \cos \eta \cos i\psi$$

$$X^{4'} = \bar{\mathbf{R}}^{4'} - \mathbf{R}^{4'} = \bar{\mathbf{R}} \cos \eta_g \sin i\psi \qquad - \mathbf{R} \cos \eta \, \sin i\psi$$

which is the ansatz for a double-surface theory. The second column on the right side gives raise to the line element

$$ds^2 = d\mathbf{R}^2 + \mathbf{R}^2 d\eta^2 + \mathbf{R}^2 \sin^2 \eta d\theta^2 + \mathbf{R}^2 \sin^2 \eta \sin^2 \theta d\varphi^2 + \mathbf{R}^2 \cos^2 \eta di\psi^2 \quad (4.4)$$

which is the line element of a pseudo-sphere for $\mathbf{R} = $ const.

   We do not care for the question, if the extra dimension has a physical reality. We use the five-dimensional ansatz as a tool for finding or explaining gravitational models. Our proposed Kerr interior results from those techniques. From (4.3) we could derive the field equations of the seed metric but we do not repeat all that we have already done to investigate the two Schwarzschild models and the Kerr exterior model. The interested reader is referred to papers [3-6]. We only give a short review how to gain the interior Kerr geometry from the pseudo-spheres. The projectors $\mathbf{P}$ operate on the fundamental quantities of the geometry:

$$\partial_a = \mathbf{P}_a^{\ b} \frac{\partial}{\partial \mathbf{R}^b}, \quad d\mathbf{R}^b = \mathbf{P}_a^{\ b} dx^a, \quad Y_{ab}^{\ \ c} = \mathbf{P}_a^{\ d} \mathbf{R}_{db}^{\ \ c} \qquad (4.5)$$

where the $\mathbf{R}_{db}^{\ \ c}$ are the connexion coefficients of the metric (4.4), the $Y_{ab}^{\ \ c}$ are the connexion coefficients of the Kerr interior surface, and

$$d\mathbf{R}^a = -\{d\mathbf{R}, \ \mathbf{R}d\eta, \ \mathbf{R}\sin\eta d\theta, \ \mathbf{R}\sin\eta\sin\theta d\varphi, \ \mathbf{R}\cos\eta di\psi\} \qquad (4.6)$$

are the spherical co-ordinate differentials. The projected surface we call the *physical surface*. The components of the projectors are

$$\mathbf{P}_0^0 = \alpha_R, \quad \mathbf{P}_1^1 = \alpha_R$$

$$\mathbf{P}_0^2 = \frac{R}{\rho_H} \sin^2 \eta, \quad \mathbf{P}_1^2 = \frac{R}{\rho_H} \sin\eta\cos\eta, \quad \mathbf{P}_2^2 = -\frac{R}{\rho_E}\sin\eta \ . \tag{4.7}$$

$$\mathbf{P}_3^3 = -\frac{R}{\rho_E} a_R^2 \sin\eta, \quad \mathbf{P}_4^4 = -\frac{\cos\eta}{\left(1 + 2\Phi_g^2\right)\cos\eta_g - \cos\eta}\frac{1}{a_R}$$

From the Riemann tensor in the five-dimensional flat embedding space

$$R_{abc}^{\ \ d}(\mathbf{R}) = 2\Big[ \mathbf{R}_{[b\cdot c\cdot\,,a]}^{\quad\ \ d} + \mathbf{R}_{[b\cdot c\cdot}^{\quad f}\mathbf{R}_{a]f}^{\ \ d} + \mathbf{R}_{[ba]}^{\quad f}\mathbf{R}_{fc}^{\ \ d} \Big] \equiv 0, \quad \Phi_{,a} = \frac{\partial\Phi}{\partial X^a} \tag{4.8}$$

we get by projection

$$\mathbf{P}_a^g \mathbf{P}_b^h R_{ghc}^{\ \ d}(\mathbf{R}) = R_{abc}^{\ \ d}(Y)$$

$$R_{abc}^{\ \ d}(Y) = 2\Big[ Y_{[b\cdot c\cdot|a]}^{\quad\ d} + Y_{[b\cdot c\cdot}^{\quad f}Y_{a]f}^{\ \ d} + Y_{[ba]}^{\quad f}Y_{fc}^{\ \ d} + Y_{gc}^{\ \ d}\left(\mathbf{P}_f^g\right)^{-1}\mathbf{P}_{[a\,\|\,b]}^{\ \ f} \Big] = 0 \tag{4.9}$$

the Riemann of the physical surface. Shifting all components with 0-indices to the right we obtain on the left the Riemann for the seed metric. Contracting to the Ricci we can construct on the right side of the Einstein field equations the stress-energy tensor, consisting mainly of the generalized second fundamental forms of the surface

$$Y_{10}^{\ 1} = M_0 = -\frac{1}{Ra_R}, \quad Y_{20}^{\ 2} = B_0 = \frac{1}{\rho_E}\sin\eta$$

$$Y_{30}^{\ 3} = C_0 = \frac{1}{\sigma}\sin\eta\sin\theta, \quad Y_{40}^{\ 4} = -E_0 = \frac{1}{\rho_g a_R a_T}\cos\eta \ . \tag{4.10}$$

$$a_T = \frac{1}{2}\Big[\left(1 + 2\Phi_g^2\right)\cos\eta_g - \cos\eta\Big]\Phi_g^{-2}$$

At the end, we obtain the components of the covariantly conserved stress-energy tensor listed in the preceding paper. How to gain the field equations and the stress-energy tensor for the rotating interior metric has been treated in this paper in full length.

## 5. Outlook

In the last Section we have briefly outlined the possibility to formulate the interior Kerr model in terms of five-dimensional differential geometry. We expect the equations for the curvatures to decouple from the Einstein field equations and the equations for the dynamical quantities to take a simpler structure. We hope to publish this elsewhere.